\documentclass[5p]{elsarticle}

\usepackage[usenames]{color}
\definecolor{darkblue}{rgb}{0,0.08,0.45}
\usepackage[bookmarksopen=true,bookmarksopenlevel=2,colorlinks=true,linkcolor=darkblue,anchorcolor=darkblue,citecolor=darkblue,urlcolor=darkblue,pdfstartview=FitH]{hyperref}
\usepackage[all]{hypcap}
\usepackage{listings}
\usepackage{booktabs}
\usepackage{amsmath}
\usepackage{multirow}
\usepackage{graphicx}
\usepackage[position=top]{subcaption}
\usepackage{pbox}

\definecolor{dkgreen}{rgb}{0,0.6,0}
\definecolor{gray}{rgb}{0.5,0.5,0.5}
\definecolor{mauve}{rgb}{0.58,0,0.82}

\lstdefinelanguage{MyPython}[]{python}
{
  numbersep=5pt,                  
  backgroundcolor=\color{white},      
  showspaces=false,               
  showtabs=false,                 
  rulecolor=\color{black},        
  tabsize=2,                      
  captionpos=b,                   
  breaklines=true,                
  breakatwhitespace=false,        
  keywordstyle=\bfseries,          
  commentstyle=\color{gray},       
  stringstyle=\color{mauve},         
  numberstyle=\bfseries
  escapeinside={\%*}{*)},            
}

\begin{document}


\title{Scalable Metropolis Monte Carlo for simulation of hard shapes}
\author[che]{Joshua~A.~Anderson}

\author[mse]{M.~Eric~Irrgang}

\author[che,mse]{Sharon~C.~Glotzer\corref{cor1}}
\ead{sglotzer@umich.edu}

\cortext[cor1]{Corresponding author}
\address[che]{Department of Chemical Engineering, University of Michigan, 2800 Plymouth Rd. Ann Arbor, MI 48109, USA}
\address[mse]{Department of Materials Science and Engineering, University of Michigan, 2300 Hayward St. Ann Arbor, MI 48109,USA}


\begin{abstract}

We design and implement HPMC, a scalable hard particle Monte Carlo simulation toolkit, and release it open source as part of HOOMD-blue.
HPMC runs in parallel on many CPUs and many GPUs using domain decomposition.
We employ BVH trees instead of cell lists on the CPU for fast performance, especially with large particle size disparity, and optimize inner loops with SIMD vector intrinsics on the CPU.
Our GPU kernel proposes many trial moves in parallel on a checkerboard and uses a block-level queue to redistribute work among threads and avoid divergence.
HPMC supports a wide variety of shape classes, including spheres / disks, unions of spheres, convex polygons, convex spheropolygons, concave polygons, ellipsoids / ellipses, convex polyhedra, convex spheropolyhedra, spheres cut by planes, and concave polyhedra.
NVT and NPT ensembles can be run in 2D or 3D triclinic boxes.
Additional integration schemes permit Frenkel-Ladd free energy computations and implicit depletant simulations.
In a benchmark system of a fluid of 4096 pentagons, HPMC performs 10 million sweeps in 10 minutes on 96 CPU cores on XSEDE Comet.
The same simulation would take 7.6 hours in serial.
HPMC also scales to large system sizes, and the same benchmark with 16.8 million particles runs in 1.4 hours on 2048 GPUs on OLCF Titan.

\end{abstract}


\maketitle{}


\section{Introduction}

CPU performance hit a performance brick wall in 2005~\cite{Asanovic2006}, and \emph{serial} execution performance has remained stagnant since then.
Whole socket CPU performance continues to increase due to additional CPU cores and wider single instruction multiple data (SIMD) vector instruction widths.
Moore's Law drives the increase in core counts, with 2-core CPUs available in 2005 increasing to 18-core CPUs in 2015.
XSEDE~\cite{Towns2014} Comet is a modern commodity dual-socket CPU cluster with 24 cores per node (12 cores per CPU).
This is a typical configuration for current systems; future clusters will have more cores per node.
However, CPUs are not very power efficient.
Graphics processing units (GPUs) have thousands of cores and can process hundreds of thousands of concurrent lightweight threads.
Given a fixed power budget, systems that use GPUs provide significantly higher performance than those with CPUs alone.
For example, the GPUs on OLCF Titan provide over 90\% of its performance. When Jaguar was upgraded to Titan by adding GPUs, its total peak performance increased by 10x with only a 20\% increase in power usage.

Metropolis Monte Carlo (MC) simulations for off-lattice particles are usually implemented in serial.
This is the most straightforward way to evolve the Markov chain, but it can achieve only a small fraction of the performance available in a single compute node and does not scale to simulations with large numbers of particles.
Efficient sampling algorithms can achieve orders of magnitude better performance than Metropolis MC, such as the event chain algorithm for hard spheres~\cite{Bernard2009a} and general pair potentials~\cite{Michel2014a}.
However, it is not clear how event chain MC can be extended to hard particles with shape, which is our primary interest.

Computational scientists need general purpose simulation tools that utilize parallel CPUs and GPUs effectively.
They need to run simulations of a few thousand particles as fast as possible in order to answer research questions quickly, conduct high throughput screening studies, and sample more states with a short turnaround time.
Researchers also need scalable codes to complete large simulations with millions of particles, which is intenable with a serial code.
There are a number of possible routes to parallelizing Metropolis MC~\cite{Heffelfinger2000}: 1) execute many independent runs in parallel to improve sampling, 2) evaluate energies in parallel for trial moves that are proposed in serial, and 3) propose multiple trial moves in parallel.
Executing many independent serial runs is not helpful for large systems or those with long equilibration times.
Two recent open source codes fall into the second category. CASSANDRA~\cite{Shah2011} uses OpenMP to run in parallel on the CPU and GOMC~\cite{Mick2013} uses CUDA to parallelize on NVIDIA GPUs.
Both of these tools model atomistic systems with classical potentials.

A number of works use checkerboard techniques to propose trial moves in parallel in off-lattice systems with short ranged interactions~\cite{Heffelfinger1996a,Uhlherr2002a,Ren2007a,OKeeffe2009a,Anderson2013}.
Heffelfinger introduced the concept~\cite{Heffelfinger1996a}, but found it inefficient due to high communication overhead.
Ren~\cite{Ren2007a} and O'Keeffe~\cite{OKeeffe2009a} improved efficiency with sequential moves in the domains.
Sequential moves obey balance in serial implementations~\cite{Manousiouthakis1999}, but the same argument does not apply to checkerboard parallel moves.
We showed in Ref.~\cite{Anderson2013} that sequential moves within active checkerboard domains lead to incorrect results, as does allowing particle displacements to cross from an active domain to an inactive one as allowed in refs~\cite{Heffelfinger1996a,Ren2007a,OKeeffe2009a}.
Uhlherr~\cite{Uhlherr2002a} implemented a two color asymmetric striped decomposition, proposes complex polymer conformation moves within the domains, and correctly rejected moves that cross boundaries.
Kampmann~\cite{Kampmann2015} combined event chain MC with the parallel checkerboard scheme in a rejection free manner by reflecting trial moves off the domain walls.

Previously, we developed a general algorithm for massively parallel Metropolis Monte Carlo, implemented it for two-dimensional hard disks on the GPU~\cite{Anderson2013}, and used it to confirm the existence of the hexatic phase in hard disks~\cite{Engel2013}.
In this paper, we present a general purpose code for MC simulations of hard shapes, HPMC.
HPMC runs NVT and NPT~\cite{Frenkel2002,Brumby2011} ensembles in 2D or 3D triclinic boxes.
Additional integration schemes permit Frenkel-Ladd~\cite{Frenkel2002} free energy computations and implicit depletant simulations~\cite{Glaser2015d}.
It calculates pressure in NVT simulations by volume perturbation techniques~\cite{Eppenga1984,Brumby2011}.
HPMC supports a wide variety of shape classes, including spheres / disks, unions of spheres, convex polygons, convex spheropolygons, concave polygons, ellipsoids / ellipses, convex polyhedra, convex spheropolyhedra, spheres cut by planes, and concave polyhedra.
It runs efficiently in serial, on many CPU cores, on a single GPU, and on multiple GPUs.
Researchers have already used HPMC in studies of shape allophiles~\cite{Harper2015} and ellipsoids with depletants~\cite{Hsiao2015a}.

\begin{figure}
\begin{lstlisting}[language=MyPython]
import hoomd_script as hoomd
from hoomd_plugins import hpmc

# Read the initial condition.
hoomd.init.read_xml(filename='init.xml')

# MC integration of squares
mc=hpmc.integrate.convex_polygon(seed=10,
                           d=0.25, a=0.3)
square=[(-0.5, -0.5), ( 0.5, -0.5),
        ( 0.5,  0.5), (-0.5,  0.5)])
mc.shape_param.set('A', vertices=square)

# Run the simulation
hoomd.run(10e3)
\end{lstlisting}
\caption{\label{fig:example_script} Example HPMC job script. The syntax is preliminary and may change as we reorganize components for a final release.}
\end{figure}

\section{Implementation}

HPMC is an extension of HOOMD-blue~\cite{Anderson2008a,Glaser2014c,HOOMDWeb} using the existing file formats, data structures, scripting engine, and communication algorithms.
HOOMD-blue started off as a molecular dynamics (MD) package, but its design is general enough to allow the addition of Monte Carlo moves with minimal modifications.
HPMC is an Integrator class inside HOOMD-blue that applies MC trial moves to the particles.
The code is object-oriented and extensible, and it is easy to add additional shape classes and collective moves.
Adding new types of local moves is not as easy, but can be accomplished by subclassing the integrator and re-implementing the main loop.

Python job scripts control HOOMD-blue execution.
Users can activate HPMC integration with a few lines, and can switch back and forth between MC and MD in the same job script.
\autoref{fig:example_script} shows a  job script that runs a simulation of hard squares for ten thousand steps.
A single ``step'' in HPMC is approximately $n_\mathrm{s}$ sweeps, the approximation is due to the parallel domain decomposition.
One sweep is defined as $N$ trial moves, where $N$ is the number of particles in the simulation box.

\subsection{Metropolis Monte Carlo}

Hard particle simulations have infinite potential energy when any particles overlap and zero potential energy otherwise.
Metropolis Monte Carlo~\cite{Metropolis1953,Frenkel2002} for hard particles with shape consists of the following steps.
Let $\vec{r}_i$ and $q_i$ be the position and orientation of particle $i$.
\begin{enumerate}
\item Select a particle $i$ at random.
\item Generate a small random trial move for that particle, resulting in a new trial configuration $\vec{r}_\mathrm{trial} = \vec{r}_i + \delta \vec{r}$, $q_\mathrm{trial} = q_i \cdot \delta q$.
\item Check for overlaps between the trial configuration and all other particles in the system.
\item Reject the trial move if there are overlaps, otherwise accept the move and set $\vec{r}_i \leftarrow \vec{r}_\mathrm{trial}$, $\vec{q}_i \leftarrow \vec{q}_\mathrm{trial}$.
\end{enumerate}

The last step is a simplification of the more general Metropolis acceptance criterion~\cite{Metropolis1953} for hard particle systems.
It offers an important opportunity for optimization: Once the first overlap is found, no further checks need to be made.

For new simulations, we follow a general rule of thumb and select the size of $\delta\vec{r}$ and $\delta q$ so that an (estimated) optimal percentage of the trial moves are accepted.
A simple way to measure efficiency for fluids is the diffusion rate in wall clock time units.
We check with this metric for several benchmark cases, and trial move sizes associated with a 20\% acceptance ratio are at or very close to peak efficiency for the high density fluids we are interested in.
The rule of thumb is not always optimal, but it is useful as researchers can trivially implement it.
All benchmark results reported in this work are initially tuned to 20\% acceptance, then the trial move size is fixed.

\subsection{Acceleration structures}

A na\"{i}ve implementation of hard particle MC would check $N-1$ particles for possible overlaps with each trial configuration.
The cost of a single sweep would be prohibitively slow: $O(N^2)$.
Acceleration structures are data structures that reduce the execution time by efficiently identifying a subset of the $N$ particles that possibly overlap with the trial configuration.
Cell lists place particles in cells and have constant lookup time to find possible overlaps: $O(N)$ sweep execution time.
Bounding volume hierarchies (BVH) build a binary tree of nodes that contain particles and have logarithmic lookup time: $O(N \log(N))$ sweep execution time.
HPMC uses cell lists on the GPU and BVHs on the CPU.

Cell lists are applicable on the CPU as well, but we were able to optimize the BVH code to be faster in all test cases.
The $\log(N)$ factor is always small as the number of particles per rank is never very large in domain decomposition parallel runs.
HOOMD-blue can compute BVHs on the GPU and use them for MD simulations as well, the focus of another publication~\cite{Howard2015}.
In this work, we do not attempt to use BVHs on the GPU as they are most useful in simulations with large particle size disparity where there is inherently very little parallelism for the GPU implementation to utilize.

\subsubsection{Cell list}

Let $d_i$ be the diameter of the sphere that encloses particle $i$, centered on the position of the particle $\vec{r}_i$.
Let $d_\mathrm{max} = \max( \left\{ d_i \right\} )$ be the largest diameter in the system.
A cell list~\cite{Frenkel2002} data structure splits the simulation box into an $n$ by $m$ by $k$ grid such that the shortest dimension of a grid cell is greater than or equal to $d_\mathrm{max}$.
Each grid cell lists the particle indices that are inside that cell.
A bucket sort efficiently assigns particles to cells, and a massively parallel bucket sort on the GPU is trivial to implement with atomic operations.
With this data structure, the cost to find possible overlapping particles is $O(1)$.
Given the position of a trial configuration $\vec{r}_\mathrm{trial}$, only its cell and the neighboring cells $\pm 1$ in each direction need to be searched (9 in 2D, 27 in 3D).
HOOMD-blue already contains code to generate cell lists for MD simulations.

Cell lists perform well, provided that particle diameters $d_i$ are comparable and there are only a few particles per cell.
Efficiency drops precipitously with disparate particle sizes.
The cell size is set by the largest diameter, so the smaller particles are able to fill in gaps leading to hundreds of particles per cell.
At large enough diameter ratios this degrades to $O(N^2)$ sweep time.

\subsubsection{Bounding volume hierarchy}

\begin{figure}
\includegraphics[width=\columnwidth]{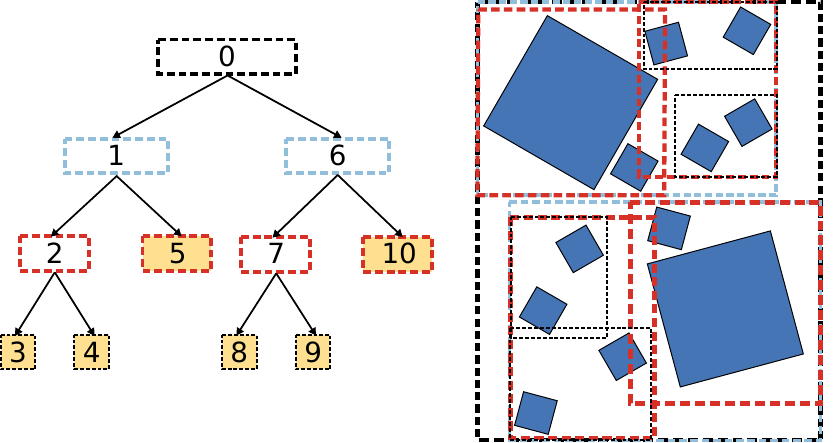}
\caption{\label{fig:aabb} A binary tree of axis aligned bounding boxes (AABBs). Each node stores an AABB that contains all of its children. Leaf nodes contain particles (2 each in this example). The tree adapts to density fluctuations and variable particle sizes. Nodes are stored in a simple array in memory, in the same order as a pre-order traversal. Numbers in the boxes indicate the index of the node in the array. Bounding volumes are visualized on the right.
}
\end{figure}

A bounding volume hierarchy (BVH) is a tree where each node represents the volume that bounds the particles inside it, as shown in \autoref{fig:aabb}.
The root node encompasses all $N$ particles, its children each encompass disjoint subsets of the particles, and so on recursively down to the leaf nodes which contain particles but have no children.
We implement binary trees of axis-aligned bounding boxes (AABB) as they have a good performance to efficiency trade-off.
AABB trees are commonly used in raytracing~\cite{Goldsmith1987,MacDonald1990,Smits1998,Karras2013} and game physics engines.
Every node in the BVH adapts to the size of the particles it contains, unlike the cell list where every cell is sized by the largest particle in the entire system.
This makes BVH trees efficient even for large size ratios.

The cost to find particles that may overlap with with a trial configuration is $O(\log(N))$.
First, compute the AABB of the trial configuration, $\mathbf{B}_\mathrm{trial}$ centered on $\vec{r}_\mathrm{trial}$ and encompassing the circumsphere of the shape.
Start searching at the root node of the tree.
At each node, check the left and right children to see if their AABB overlaps with $\mathbf{B}_\mathrm{trial}$.
If a child does not overlap, skip it and all of its children.
If it does overlap, recurse down and check its children.
If this is a leaf node, check the trial configuration for overlaps against all particles in the leaf.
When a trial move is accepted, update the tree so that future moves will not miss possible overlaps.
We update the tree without changing its topology by expanding the AABB of the leaf and all of its parent nodes recursively in $O(\log(N))$ steps.

The main simulation loop performs trial moves, and the next innermost loop finds possible overlaps.
Recursive function calls have high overhead, and an explicit stack is almost as expensive.
Neither performs well in the inner loop.
We use a stack-free iterative scheme to perform the search, and implement AABB overlap tests using streaming SIMD extensions (SSE) vector intrinsics.
We store the nodes in an array in pre-order and include an additional skip count for each node.
With this structure, the recursive algorithm becomes a simple \emph{for} loop iteration over nodes in memory order~\cite{Smits1998}.
When a node and all of its children need to be skipped, add the pre-computed skip value to the loop index.
Iterating over nodes in memory order is cache-friendly on the CPU and gives good performance even with large $N$.

Unlike cell lists, AABB trees do not directly encode periodic boundary information.
When a particle is near a boundary, or in a small simulation box where particles may interact with themselves, HPMC translates each trial configuration $\vec{r}_\mathrm{trial}$ by all necessary periodic images and checks each image against the tree separately.
Most of these checks terminate the search at the root node, so there is little performance penalty.
There are no periodic boundaries in a single domain of a fully decomposed parallel simulation, so only one tree search is needed in the most common use-case.

Particles migrate to and from domains, and it would be impossible to maintain a balanced tree with constant additions and removals.
So at the start of each step, we destroy the old AABB tree and build a new one from scratch.
There are many tree build algorithms, including top-down and bottom up approaches.
There are different heuristics for determining node splits~\cite{MacDonald1990}, and methods to optimize trees after they are built~\cite{Karras2013}.
Low quality trees are fast to build but take longer to traverse in the search phase.
High quality trees cost more execution time to construct, but are faster to traverse.

We choose the median cut algorithm~\cite{Goldsmith1987} as it is simple to implement and offers good tree quality and very fast build times.
It proceeds as follows:
For each particle, construct the AABB that bounds the particle's circumsphere.
Partition that list recursively.
At each level of recursion, merge the current sublist of AABBs into one large AABB.
Split that AABB at the median of its longest axis.
Partition the particles in the sublist so that particle centers less than or equal to the split are on the left side of the array and particle centers greater than the split are on the right.
Terminate recursion and generate a leaf node when the number of particles in the sublist is less than or equal to the maximum node capacity (we use 12, which we found empirically).
If recursion is not terminated, generate a new internal node with the merged AABB and place it at the end of the array.
This build algorithm puts the nodes in pre-order to be searched with the stack-free iterative scheme.

The entire recursion partitions a single array of AABBs and requires no memory allocations or extra copies of data.
A na\"{i}ve implementation with memory allocations and data copies takes many times longer to execute.
Tighter AABBs are possible for shaped particles, but are much more expensive to compute as they need to be updated every time the particle rotates.
With an AABB that bounds the circumsphere, the particle can rotate without changing its AABB and this reduces the time needed to update the tree on an accepted trial move at the cost of a slightly lower quality tree.

\subsection{Overlap checks}

HPMC supports many different classes of shapes.
It calls the shape overlap check from the innermost loop and executes it billions of times per second in a typical simulation run, so heavily optimized shape overlap checks are needed for good overall performance.
We write each shape overlap check ourselves and do not use existing libraries that would require costly data conversions and function call overhead at every check, and which lack GPU support.
There is a single MC integration loop that is templated on the shape class to enable the best performance and to make code maintenance easy.
Only that single class needs to be modified when fixing bugs or adding additional features to the main loop.
By template instantiation, the compiler is able to inline every overlap check call, we can arrange the data in the best format for the computation, and we can choose the best overlap detection algorithm for each shape class.
Users only need to write an overlap check to add a new shape class.

There are a number of methods to determine if two shapes overlap.
Some methods are specific for a single class of shapes, while other are more general.
When there are multiple algorithms to choose from, we test them and select the one with the best performance.
For spheres and disks, overlap detection is trivial.
For unions of spheres, all spheres in one shape are exhaustively checked against all those in the other.
We use the separating planes method~\cite{Gottschalk1996} for convex polygons on the CPU, but XenoCollide~\cite{Jacobs2008} is faster on the GPU.
XenoCollide is a general algorithm that can detect overlaps between any two convex shapes.
HPMC uses XenoCollide for convex polygons, spheropolygons, convex polyhedra, convex spheropolyhedra, and spheres cut by planes.
To detect overlaps between two concave polygons, HPMC checks all pairs of edges and all vertices.
If no edges intersect and no vertex from one shape is inside the other, then the shapes do not overlap.
We use a matrix method~\cite{Wang2001a,Alfano2003,Hsiao2015a} to detect overlaps of ellipsoids and ellipses.

It would be expensive in both memory and compute to keep all particle geometry (e.g. polyhedron vertices) in world coordinates.
HPMC efficiently represents each particle with a position $\vec{r}_i$ and an orientation quaternion $q_i$.
Together, these describe how to rotate and then translate from the frame of the particle to the world frame.
The user specifies the shape geometry in particle local coordinates once for each type of particle.

When performing an overlap check between particles $A$ and $B$, HPMC works in a local coordinate system centered on particle $A$.
The application of this local coordinate system optimization depends on the shape overlap check algorithm.
For example, a support function is evaluated at every iteration of XenoCollide.
The support function for the Minkowski difference $B-A$ is
\begin{equation}
\mathbf{S}_{B-A}^{\text{world}}(\vec{n}) = \mathbf{S}_B^{\text{world}}(\vec{n}) - \mathbf{S}_A^{\text{world}}(-\vec{n})
\;\text{.}
\end{equation}
Replacing $S_A^{\text{world}}$ and $S_B^{\text{world}}$ with operations on the particle support functions in their local coordinate systems gives~\cite{Jacobs2008}
\begin{equation}
\mathbf{S}_{B-A}(\vec{n}) = \mathbf{R} \mathbf{S}_B(\mathbf{R}^{-1} \vec{n}) + (\vec{r}_B - \vec{r}_A) - \mathbf{S}_A(-\vec{n})
\end{equation}
where $\mathbf{R}$ is the rotation matrix that takes $B$ into the coordinate system of $A$.
$\mathbf{S}_{B-A}$ is in the coordinate system of particle $A$, but this is irrelevant for the overlap calculation.
On the GPU, we replace $\mathbf{R}$ with an operation that rotates vectors by the quaternion $q=q_A^* q_B$ because it is faster.
The quaternion rotation uses more floating point operations, but requires fewer registers.

Single precision particle coordinates are not accurate in large simulation boxes (without cell-local coordinate systems~\cite{Anderson2013}), so in HPMC we use double precision particle coordinates.
In mixed precision mode, we compute the displacement between particles $\vec{r}_{AB} = \mathrm{min\_image}(\vec{r}_B - \vec{r}_A)$ in double precision, then we cast $\vec{r}_{AB}$ to single precision and compute the overlap check in single precision.
Within the local coordinate system of particle $A$, single precision is accurate for self-assembly simulations, though densest packing calculations may require full double precision.
HPMC supports both full double precision and mixed precision modes as a compile time option.
Full single precision builds do not even pass simple validation tests.

All benchmark and validation studies in this work use mixed precision.

\subsection{SIMD vectorization}

The polygon, spheropolygon, polyhedron, and spheropolyhedron overlap checks evaluate the support function many times in the innermost loop of XenoCollide.
The support function loops over all vertices in the shape, dots them with $\vec{n}$, and returns the vertex that gives the maximum dot product.
In our initial implementation, this code used over 80\% of the CPU time (determined by line level profiling with \texttt{oprofile}).
We improve performance of this loop with SIMD vector instruction intrinsics.
The first loop computes the dot products for all vertices, $w$ vertices per iteration with SIMD parallelism,
and just stores the result to avoid branch mispredication penalties around the floating point operations.
A second $w$ width SIMD loop starts and each iteration uses masks and the \texttt{BSF} assembly instruction to find the index of the maximum element.
We implement these loops in SSE ($w=4$) and AVX (advanced vector extensions) ($w=8$).
SIMD vectorization boosts performance of the support function evaluation by a factor of 2-3 over a serial implementation with manually unrolled loops, achieving near peak floating point throughput in a microbenchmark.
While the vectorized support function now executes several times faster, it is only one part of a production simulation run.
We used \texttt{oprofile} to run a line level execution profile of a typical polyhedra simulation in the final version of the code.
About 40 percent of the runtime is spent in the vectorized support function, 10 percent in XenoCollide iteration logic, 40 percent in AABB tree searches and the remaining 10 percent in trial moves and AABB tree generation.

\subsection{Parallelization}

\begin{figure}
\includegraphics[width=\columnwidth]{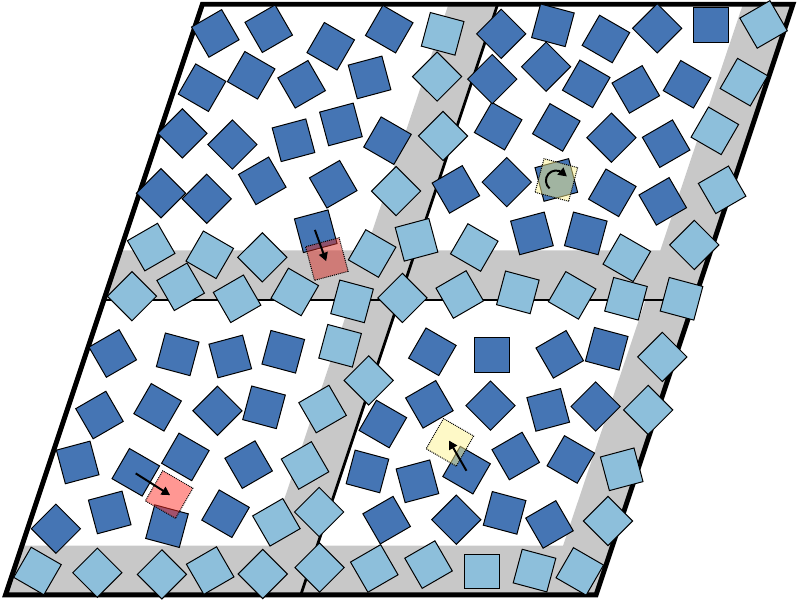}
\caption{\label{fig:parallelization} Domain decomposition scheme. The outer box is the triclinic simulation box, which is split into 4 domains. On the right and bottom edge of each domain is a gray inactive area, one particle diameter wide. Particles in the inactive region are colored lighter and are not selected for trial moves. Any trial configuration that ends in the inactive region must be rejected (top left domain in this example). On the GPU, individual domains are further subdivided with a checkerboard grid.}
\end{figure}

Even with fast BVH trees and SIMD vector optimizations, a serial CPU simulation still only uses a fraction of the capabilities of a single compute node.
We implement parallel computations that utilize the full capabilities of multi-core CPUs and clusters of CPU nodes to provide faster time to solution and to enable larger scale simulations across many nodes.
In hard particle MC, there are typically only a few dozen possible overlaps with each trial configuration.
This is not large enough to parallelize over a whole node and cannot scale to large simulations.
The only path to achieving fast, scalable simulations for MC with short range particle interactions is to perform many trial moves in parallel~\cite{Heffelfinger2000}.

To do this, we need to be able to efficiently generate many parallel random number streams.
As we have before~\cite{Phillips2011, Anderson2013}, we use a hash based RNG, Saru~\cite{Afshar2013}.
Each time a trial move is generated, we hash together the particle index, time step, user seed, and MPI rank to initialize an independent RNG stream.
We then use that stream to generate as many random numbers as needed for the trial move.

\subsubsection{Domain decomposition}

To scale beyond a single compute node, we employ a domain decomposition strategy using MPI with one rank per CPU core, or one rank per GPU.
We implement HPMC as an extension of HOOMD-blue, a parallel MD code that already has the necessary decomposition and communications routines~\cite{Glaser2014c}.
Each rank covers a portion of the simulation box and owns all of the particles in that region.
The communications routines copy particle data from neighboring ranks in a ghost layer around each domain, and migrate particles from one domain to another as they move.

We base HPMC domain decomposition on our previous method for massive parallelism~\cite{Anderson2013}.
However, we do not use a $2^d$ color checkerboard grid to scale across domains.
Updating only $\frac{1}{2^d}$ of the system at a time is unnecessary with low thread counts, and would require ghost communication after every fractional system update.
Instead, we modify the checkerboard scheme to have only two regions (active and inactive) and make the active region as large as possible, see \autoref{fig:parallelization}.
The inactive region has width $d_\mathrm{max}$, and it is placed along the bottom, right, and back faces of each domain.

In this layout, all inactive particles are in the neighboring domain's ghost layer, or separated from the neighboring domain's active particles by an inactive region.
There is no need to communicate ghost layer updates because these particles do not move during substeps.
Communication between ranks only occurs at the end of the step, when we apply a single random displacement to all particles and call the migration routine.
The user sets $n_\mathrm{s}$, the number of substeps to perform per step, giving them control of the computation to communication ratio.

In our previous work~\cite{Anderson2013}, we showed that shuffling the order of particles selected for trial moves achieves detailed balance within the checkerboard scheme.
We proposed full shuffling of all 0--4 particle indices within a cell, as forward and reverse permutations occur with equal probability.
Full shuffling causes cache thrashing in a general CPU domain decomposition implementation, where individual domains might have thousands of particles.
In HPMC, we choose randomly to loop through particles either in forward or reverse index order.
Both orders are cache friendly, and this selection preserves the essential element required for detailed balance: that forward and reverse sequences occur with equal probability.
With this slight modification, this scheme obeys detailed balance following the same arguments as in ref.~\cite{Anderson2013}.

Putting all of these elements together, HPMC with domain decomposition on the CPU has the following stages.
\begin{enumerate}
\item Generate the AABB tree.
\item Choose forward or reverse index order randomly.
\item Loop through all particles $i$ in the chosen order, skipping those where $\vec{r}_i$ is in an inactive region.
\item Generate a small random trial move for particle $i$, resulting in a new trial configuration $\vec{r}_\mathrm{trial} = \vec{r}_i + \delta \vec{r}$, $q_\mathrm{trial} = q_i \cdot \delta q$.
Reject the trial move if $\vec{r}_\mathrm{trial}$ is in an inactive region.
\item Check for overlaps between the trial configuration and all other particles in the system, using the AABB tree.
\item Reject the trial move if there are overlaps, otherwise accept the move and set $\vec{r}_i \leftarrow \vec{r}_\mathrm{trial}$, $\vec{q}_i \leftarrow \vec{q}_\mathrm{trial}$.
Also, update the AABB tree with the new position of particle $i$ which may or may not require expanding its leaf node and all parents up to the root.
\item Repeat stages 2--6 $n_\mathrm{s}$ times.
\item Choose a random displacement vector and translate all particles by this vector.
\item Migrate particles to new domains and communicate ghost particles.
\end{enumerate}

Stages 1--9 implement one \emph{step}, and typical MC simulation runs continue for tens of millions of steps.
The amount of useful work done by a step is proportional to the number of trial moves attempted and simulation effort is usually measured in sweeps ($N$ trial moves).
When running on a single rank, one step executes $n_\mathrm{s}$ sweeps.
The ratio of active to inactive particles decreases as the number of parallel domains increases, so the number of sweeps in a step varies depending on the run configuration.
Users need to be aware of this behavior so that they can configure their run protocols properly.

\subsubsection{GPU kernel}

\begin{figure*}
\center
\includegraphics{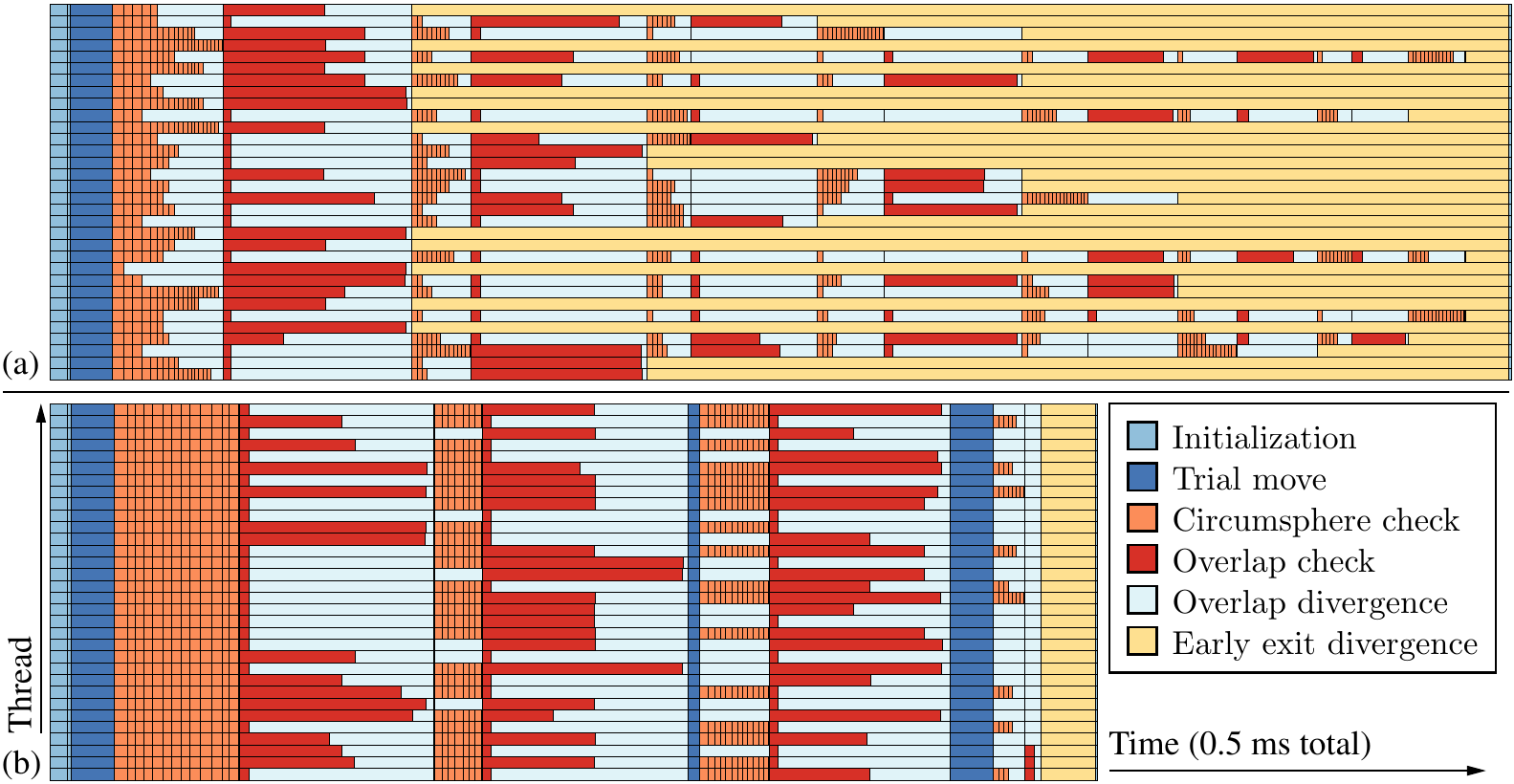}
\caption{\label{fig:queue} Traces of warp execution from a benchmark run of truncated octahedra. Timing data is captured with the \texttt{clock64()} function for just a single warp on the device. Colors indicate time spent in different parts of the execution. Panel (a) shows the register queue implementation and (b) shows the block queue. In panel (b), the later trial move and early exit condition rectangles indicate synchronization with other warps in the block.}
\end{figure*}

For multi-GPU simulations, we use the same domain decomposition strategy as on the CPU but assign each active domain to a single GPU.
On the GPU, we run a kernel that implements the checkerboard update scheme, similar to the one we previously implemented~\cite{Anderson2013} but with a few differences.
In HPMC, user configuration choices can lead to hundreds of particles in a cell, so we keep particle positions in global memory and each kernel call only proposes one trial move per cell.
For disk simulations, this is slower than our specialized implementation~\cite{Anderson2013}, but it is not a bottleneck for complex shaped particles where the overlap check costs dominate and accessing global memory is almost free in comparison.

To assign threads to cells, we pre-compute arrays that list the active cells for each color of the checkerboard.
Then we launch 1D indexed kernels that read their cell from this array so that a single kernel may work for all use-cases.
Most research relevant simulations are dense enough that the fraction of empty cells is small, though these could be removed from the list with an additional overhead per step.
This structure makes one trial move for each cell that has a non-zero number of particles in it.
To approach parity between a step on the GPU and a step on the CPU, HPMC uses particle density and the number of cells to estimate how many times to run the kernel so that one GPU step is approximately $n_\mathrm{s}$ sweeps.

\subsubsection{Block queues}

For complex shaped particles, such as polyhedra with many vertices, overlap checks take a majority of the kernel run time.
They are compute limited, so GPUs have the potential to execute these checks with very high performance, but divergence is a problem.
Current GPUs execute warps of 32 threads in lockstep.
Threads within a warp can take different branches, but all threads in a warp execute the instructions on both sides of the branch and inactive threads are masked out.
A direct translation of the MPMC checkerboard algorithm~\cite{Anderson2013} loops over particles in nearby cells, checks for circumspheres that overlap with the trial configuration, and calls the full overlap check if the circumspheres overlap.
In a typical simulation, there might be 100 particles in the cells around the location of a trial configuration, but only five of those pass the circumsphere test.
With such low hit probabilities, that branch is likely to diverge every time, leading to a large reduction in performance.

We improve on this by changing the structure of the loop to make a register queue.
Threads loop over potential neighbors, only checking the circumsphere overlap inside the loop.
When a thread finds a potential overlap, it breaks out of the loop.
Then the full overlap check is performed outside the loop after the threads have converged.
This modification causes the overlap checks to run as converged as possible.
\autoref{fig:queue}(a) shows a trace from a warp using the register queue.
The next problem is immediately obvious in this figure: 80 percent of the threads end early when they find their first overlap and know that the move must be rejected.
The remaining 20 percent must check all potential overlaps before accepting the move.
The critical path for the entire warp to complete is determined by only 20\% of the threads so divergence is still a problem.

We attempt to use a global queue to work around this.
The first kernel generates trial moves, performs circumsphere checks, and inserts the needed full overlap checks into a global queue.
Then a second kernel processes the queue and runs all of the overlap checks with no divergence due to circumsphere checks or early exit conditions.
A third kernel applies the accepted moves.
Overall, this method performs no better than the register queue kernel.
It was able to compute many more overlap checks per second, but it also had to perform many more overlap checks because it is not able to take advantage of the early exit condition.

Our fastest, and final, implementation uses the idea of a work queue for the expensive overlap checks, but does so at a block level rather than at the global level.
One or more threads in a group run for each cell in the active set.
They generate the trial move and then loop through the particles in the nearby cells in a strided fashion.
For example, with a group size of 4, thread 0 checks nearby particles 0, 4, 8, \ldots and thread 1 checks 1, 5, 9, \ldots .
In this phase, threads only check for circumsphere overlaps.
When a particle passes the circumsphere test, the thread adds the particle index and group id to a queue in shared memory.
The maximum queue size is the number of threads in the block.
Once the queue is full, the loop over nearby particles exits and all threads in the block enter the next phase.
Here, each thread performs the overlap check in the queue entry matching its thread index, which may be for a trial move generated by a different thread.
If the particles overlap, the thread atomically increments an overlap counter for the appropriate group.
Then the first phase starts populating the queue again, except that threads with already discovered overlaps do not add any work to the queue.
These two phases repeat until there are no more nearby particles to check for any thread in the block, then accepted trial moves are handled.

\autoref{fig:queue}(b) shows a trace from a warp using the block queue.
The overlap check phase of the kernel runtime is kept dense and non-divergent until the last pass of the non-full queue.
There is still divergence within the iterative XenoCollide overlap checks themselves.
We implemented XenoCollide as a single loop to avoid divergence as much as possible, but some particle configurations require more iterations than others.
We tried a variety of ways to remove overlap checks that exit in the first iteration from the queue, but adding that cost on top of every circumsphere check slowed performance overall.

\begin{figure*}[t]
\centering
\subcaptionbox{Comet performance ($N=4096$)}
{
\begin{tabular}{rrrr}
\toprule
 & \multicolumn{3}{c}{Hours to run 10e6 sweeps} \\
P & Pentagon & Dodecahedron & Binary \\
\midrule
1 & 7.56 & 21.78 & 32.94 \\
2 & 3.58 & 11.33 & 17.40 \\
4 & 1.91 & 6.52 & 9.48 \\
8 & 1.06 & 3.72 & 5.86 \\
16 & 0.55 & 1.96 & 3.28 \\
24 & 0.39 & 1.37 & 2.45 \\
48 & 0.23 & 0.88 & - \\
96 & 0.17 & 0.75 & - \\
192 & 0.19 & - & - \\
384 & 0.26 & - & - \\
\bottomrule
\end{tabular}
}
\subcaptionbox{Scaling on Comet}
{
\includegraphics{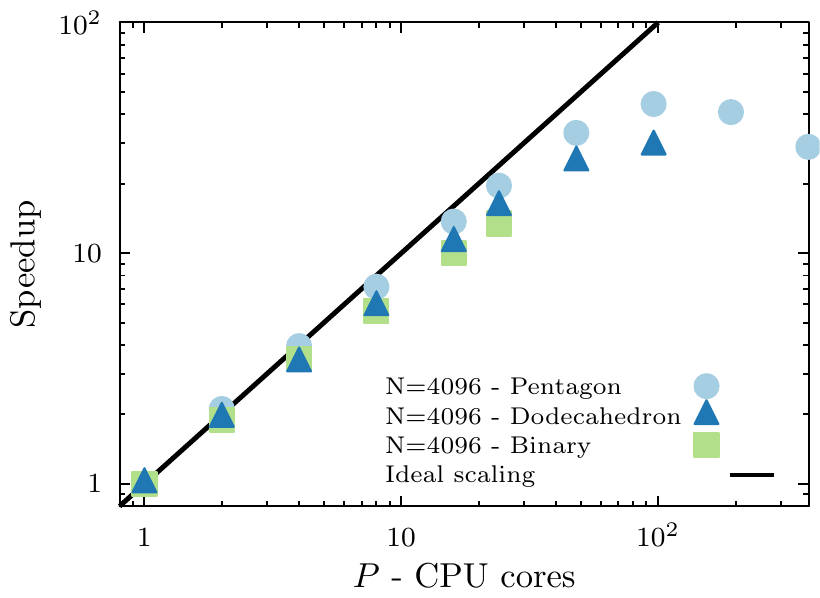}
}

\bigskip

\subcaptionbox{Titan performance (N=$2^{24}$)}
{
\begin{tabular}{rrrr}
\toprule
& \multicolumn{3}{c}{Hours to run 10e6 sweeps} \\
P & Pentagon & Dodecahedron & Binary \\
\midrule
8 & 102.03 & 492.83 & 901.73 \\
16 & 51.25 & 252.48 & 482.17 \\
32 & 26.04 & 131.01 & 294.12 \\
64 & 13.82 & 73.42 & 173.80 \\
128 & 7.47 & 44.61 & 114.48 \\
256 & 4.68 & 26.32 & 72.52 \\
512 & 2.83 & 16.64 & 47.45 \\
1024 & 1.83 & 17.13 & 32.42 \\
2048 & 1.42 & 12.40 & - \\
4096 & 1.46 & 10.36 & - \\
\bottomrule
\end{tabular}
}
\subcaptionbox{Scaling on Titan}
{
\includegraphics{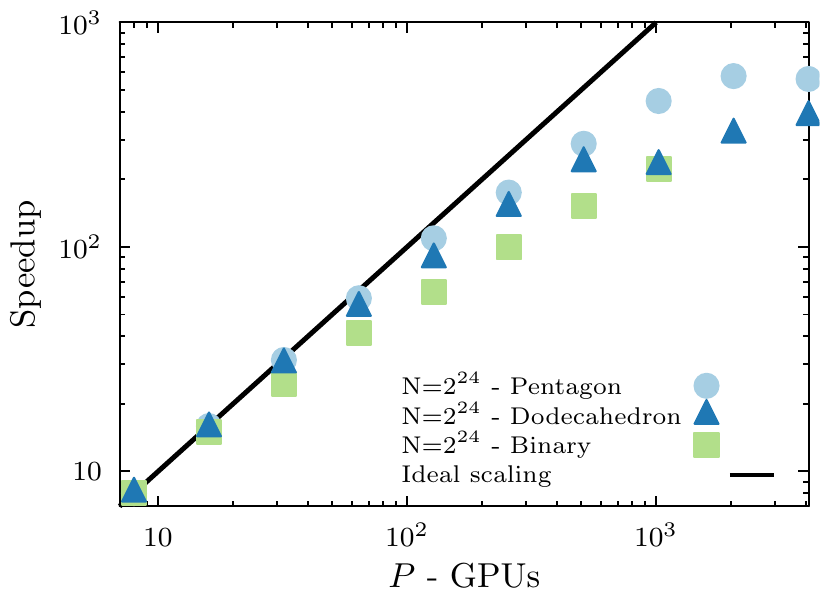}
}
\caption{\label{fig:perf} Performance is reported in hours to complete 10e6 sweeps. Scaling plots are performance values renormalized to performance on the smallest $P$ run. Simulations are run on CPUs on XSEDE Comet (Intel Xeon E5-2680v3) with $N=4096$ and GPUS on OLCF Titan (NVIDIA Tesla K20X) with $N=2^{24}$.}
\end{figure*}

With the block queue implementation, any number of threads can be run per active cell so long as the total block size is a multiple of the warp size.
This allows many threads to execute per cell, which is critical to obtain high performance on modern GPUs.
HOOMD-blue autotunes kernel launch parameters to find the fastest performing values~\cite{Glaser2014c}.
We autotune over all valid combinations of the group size and block size to find the fastest performing configuration.
In cases where there are a large number of particles in nearby cells, the autotuner will pick a large group size (i.e. 8 or 32) to have many threads available to process the overlap checks.
In cases where there are only a few particles in nearby cells, it chooses 1 or 2.
We test a variety of benchmark cases and always find the block queue outperforms the register queue; performance benefits range from 20 to 80 percent.

\section{Performance}

We benchmark HPMC performance on a few reference systems that researchers have previously studied.
Our first benchmark is a system of 2D regular pentagons in a high density fluid at a packing fraction of 0.676 in NVT.
This is a single state point in a previous study by Schilling, Frenkel et~al.~\cite{Schilling2005a}.
Our second benchmark is a system of 3D dodecahedra in a high density fluid at a packing fraction of 0.5 in NVT.
This is representative of monodisperse self-assembly simulations of polyhedra~\cite{Damasceno2012a}.
Binary systems have a much larger phase space to explore (composition, size ratio), so such studies are computationally expensive and can benefit greatly from optimized, parallel simulation codes.
Khadilkar and Escobedo~\cite{Khadilkar2012} studied a binary mixture of tetrahedra and octahedra with equal edge lengths that could tile space (volume ratio $1:4$).
We use this system for our third benchmark, in the solid at a packing fraction of 0.6 in NVT.
The binary benchmark benefits greatly from the BVH tree, though the size ratio is not large enough to demonstrate the full capabilities of the tree to efficiently simulate huge size disparities.
We leave those benchmarks for other papers on methods to model systems with large colliods and small depletants~\cite{Glaser2015d} and MD methods using BVH trees~\cite{Howard2015}.

For all three benchmarks, we explore strong scaling performance in two regimes.
The first case is $N=4096$, a system size representative of what researchers have used in previous studies with serial MC implementations.
Such systems are too small to run efficiently on the GPU, but parallel CPU simulations offer tremendous speedups over serial ones.
We run this case on XSEDE Comet, a recent addition to the XSEDE ecosystem.
Comet has dual-socket nodes with Intel Xeon E5-2680v3 CPUs --- a total of 24 cores per node.
\autoref{fig:perf}(a--b) shows the performance of the three benchmarks at $N=4096$ on Comet.
Both the pentagon and dodecahedron benchmarks scale out to 96 CPU cores, only 43 particles per domain.
At this point, it takes 10 minutes to run 10 million sweeps in the pentagon benchmark, and 45 minutes in the dodecahedron benchmark.
Contrast that with serial simulations that would take 7.6 and 21.8 hours, respectively.
Due to the size disparity, the binary benchmark does not decompose over 24 cores.
Past that point, the inactive region covers the whole domain.
Still, we reduce a serial runtime of 32.9 hours down to 2.45 to complete 10 million sweeps.

The second regime we benchmark is large systems of $N=2^{24}$ (16.8 million) particles.
Running such a large system is inconceivable with a serial simulation code, where it would take more than a month to complete 10 million sweeps and years to equilibrate a system.
Large systems easily fill the GPU, so we run these benchmarks on OLCF Titan, which has 1 NVIDIA Tesla K20X GPU per node. \autoref{fig:perf}(c--d) shows the results.
The pentagon benchmark scales out to 2048 GPUs (8192 particles/GPU), where it takes 1.4 hours to complete 10 million sweeps.
The dodecahedron benchmark scales out to 4096 GPUs (4096 particles/GPU), where it takes 10.36 hours to complete 10 million sweeps.
As on the CPU, domain size limits the scaling of the binary benchmark, this time to 1024 GPUs (16384 particles/GPU).

The strong scaling limit is the fastest possible simulation one can achieve, though it uses compute resources inefficiently.
Given a fixed compute time budget, one can get more simulations completed with fewer MPI ranks at the cost of longer wait times to finish each run.
Efficiency depends primarily on the number of particles per CPU core (or per GPU).
HPMC obtains a reasonable efficiency of 60--70\% with 85 particles per CPU core for the pentagon benchmark and 170 for the dodecahedron benchmark.
The same efficiency is reached at 65536 pentagons/GPU and 131072 dodecahedra/GPU, though Titan's usage policies strongly encourage runs much closer to the strong scaling limit.
These are representative of 2D and 3D simulations of single particle type systems in general, so researchers can use these as rules of thumb.
For systems with particle size disparity, we advise users to run their own short scaling benchmarks for their systems to determine an appropriate selection.
Efficiency as a function of $N/P$ varies greatly with size ratio and composition parameters.

GPU speedup over the CPU is not a focus of this work; we instead present what types of simulations the CPU and GPU hardware architectures are well-suited for and how well HPMC performs those benchmarks.
However, some readers may still be interested in relative speedup.
It is difficult to make a GPU/CPU comparison at scale and it is not fair to compare several year old K20X GPU to the brand new Haswell CPUs on Comet.
However, these are the systems currently available to researchers at scale --- Comet has K80s, but limits users to no more than 16 GPUs at a time.
One way to compare is to pick points at the 60--70\% efficiency level and compare trial moves per second.
At reasonable efficiency on the pentagon benchmark, HPMC performs 12.5 million trial moves per second per CPU socket and 38.9 million per GPU socket, for a socket to socket speedup of 3.1.
At reasonable efficiency on the dodecahedron benchmark, HPMC performs 4.45 million trial moves per second per CPU socket and 8.16 million per GPU socket, for a socket to socket speedup of 1.8.

\section{Validation testing}

We rigorously test HPMC for validity at three levels.
At the lowest level, we perform unit tests on the AABB tree, move generation code, and shape classes.
We test that AABBs are generated properly, and that queries on the resulting trees find all possible overlapping particles.
We verify that trial moves are generated from the proper uniform distribution and that the particle update order is correctly randomized.
For each shape class, we place many test configurations and validate that overlapping and non-overlapping configurations are correctly detected.
This is essential to ensure the quality of the overlap check algorithms as there are many corner cases to account for.
The shape overlap unit tests contain many configurations captured from simulation runs that we identified were overlapping by independent methods.
Low level unit tests cover 14 classes with over 1400 different checks.

MC integrators cannot be checked with low level unit tests because their stochastic nature makes it impossible to define what a correct output is given an input.
Instead, we validate the integrators with system level tests which are python job scripts that perform simulation runs.
System tests verify different operating modes of the integrator and ensure that documented interfaces for controlling those modes work.
Additionally, we run short simulations and check for any overlaps in the generated configurations.
Most bugs in the integrator implementation result in accepted moves that have overlapping particles.

We run unit and system level tests on every commit to the repository with a Jenkins continuous integration server.
Jenkins runs these tests on the CPU and on 4 different generations of GPU, in both mixed and double precision, and with and without MPI for a total of 20 build configurations.
It e-mails developers when a commit fails any of the tests.

Unit and system level tests are designed to run quickly and automatically to detect bugs in the implementation.
They are not sufficient to verify that HPMC correctly samples the ensemble of states available to the system.
Such tests take much longer to run and we do so by hand.
We test three separate systems to validate HPMC in NVT: disks, spheres, and truncated octahedra.
We run each system in multiple compute configurations, sample the pressure to high precision, and ensure that all simulations produce the same result.
For hard disks, we have three independent data points to compare to from event chain MC, event driven MD, and our previous GPU checkerboard implementation~\cite{Engel2013}.
No such high precision data exists to validate hard spheres and truncated octahedra, so we instead verify that serial and all parallel builds agree.
From our previous work, and from tests that introduce issues in HPMC, we know that this validation technique is sensitive enough to detect when there are subtle problems - such as looping through particles in sequence instead of randomly choosing the forward or reverse order.

\begin{table}
\centering
\begin{tabular}{rr@{.}lr@{.}lr@{.}l}
\toprule
Mode & \multicolumn{2}{c}{Disk} & \multicolumn{2}{c}{Sphere} & \multicolumn{2}{c}{\pbox{3cm}{Truncated \\ octahedron}} \\
\midrule
Serial & 9 & 1707(4) & 9 & 3135(4) & 13 & 8975(17) \\
24 CPU cores & 9 & 1709(2) & 9 & 3136(2) & 13 & 8972(11) \\
1 GPU & 9 & 1708(2) & 9 & 3134(2) & 13 & 8970(7) \\
4 GPUs & 9 & 1710(3) & 9 & 3134(2) & 13 & 8970(6) \\
\bottomrule
\end{tabular}
\caption{\label{tab:validation} Pressures obtained during NVT validation test runs, in reduced units. In 2D, $P^* = \beta P \sigma^2$, where $\sigma$ is the diameter of the disk. In 3D, $P^* = \beta P v_0$, where the reference volume $v_0$ is the single particle volume for the respective shape. The numbers in parentheses are two standard errors of the mean in the last given digit(s).}
\end{table}

Specifically, we run: 65536 hard disks at a packing fraction of 0.698 (fluid), 131072 hard spheres at a packing fraction of 0.60 (solid), and 16000 truncated octahedra at a packing fraction of 0.7 (solid).
We initialize the hard disks randomly and allow them to equilibrate, while we place the two solid systems on the known self-assembled FCC and BCC lattices.
We prepare a number of independent equilibrated initial configurations and run as many sampling runs, 30 for disks, and 8 for the other shapes.
We ran each parallel disk simulation for 60 million sweeps (only 24 million in serial), spheres for at least 8 million, and truncated octahedra for up to 80 million sweeps, all after suitable equilibration periods.
Estimated error is reported as 2 standard errors of the mean from the independent runs.
We perform the set of runs in serial, on 24 CPU cores in parallel with domain decomposition, on a single GPU with checkerboard parallelism, and on 4 GPUs with both checkerboard and domain decomposition.
Tests are performed on the XSEDE Comet and University of Michigan Flux systems.

To sample the pressure in NVT runs, we build a histogram of scale factors $s$ that cause two neighboring particles to overlap and extrapolate the probability of overlap at $s=0$.
This is a generalization of the $g(r)$ based technique we previously used for hard disks~\cite{Anderson2013,Engel2013}, see those references for full details on how to sample and extrapolate the histogram without introducing any systematic errors.
We use a more general volume perturbation technique~\cite{Eppenga1984,Brumby2011} to extend this method to particles with shape.

\autoref{tab:validation} shows the results of these tests.
For each shape, all run configurations give the same pressure within error and verify that HPMC performs correct simulations in all parallel modes.
Additionally, HPMC's hard disk results match, within error, the previous pressures obtained with three simulation methods~\cite{Engel2013}, where $P^* = 9.1707(2)$ for this state point.

\section{Conclusions}

We presented HPMC, a parallel simulation engine for hard particle Monte Carlo simulations we developed as an extension to HOOMD-blue.
HPMC executes in parallel on many CPU cores and many GPUs, and we optimized HPMC to run as fast as possible on both architectures.
On the CPU, we used efficient bounding volume hierarchies to search for possible overlaps, and SIMD vector intrinsics in the innermost loop to take full advantage of modern processors.
On the GPU, we performed trial moves in parallel on a checkerboard with many threads per cell, and implemented a block level queue to limit performance degradation due to divergence.

Our implementation is general and works for any shape, given an implementation of an overlap check.
Users can easily add new shapes to the code without needing to write GPU kernels.
HPMC ships with overlap checks for many classes of shapes, including spheres / disks, unions of spheres, convex polygons, convex spheropolygons, concave polygons, ellipsoids / ellipses, convex polyhedra, convex spheropolyhedra, spheres cut by planes, and concave polyhedra.

Completing 10 million sweeps of a system of 4096 pentagons required 7.6 hours in serial.
HPMC achieved the same in 10 minutes when running in parallel on 96 cores.
GPUs allow efficient runs with tens of millions of particles.
On 2048 GPUs, HPMC ran 10 million sweeps of a system of 16.8 million pentagons in 1.4 hours.

HPMC is available open-source in HOOMD-blue, starting with version 2.0.

\section*{Acknowledgments}

HPMC is a group effort. We acknowledge the following group members for contributions to various parts of the code, Khalid Ahmed, Michael Engel, Jens Glaser, Eric S. Harper, and Benjamin A. Schultz.

Parts of this work were supported by the DOD/ASD(R\&E) under Award No. N00244-09-1-0062 (initial design and implementation) and the National Science Foundation, Division of Materials Research Award \# DMR 1409620 (BVH tree implementation and GPU kernel optimizations).
Software was validated and benchmarked on the Extreme Science and Engineering Discovery Environment (XSEDE), which is supported by National Science Foundation grant number ACI-1053575, on resources of the Oak Ridge Leadership Computing Facility at the Oak Ridge National Laboratory, which is supported by the Office of Science of the U.S. Department of Energy under Contract No. DE-AC05-00OR22725, and was also supported through computational resources and services provided by Advanced Research Computing Technology Services at the University of Michigan, Ann Arbor.

The Glotzer Group at the University of Michigan is an NVIDA GPU Research Center. Hardware support by NVIDIA Corp. is gratefully acknowledged.

\section{Bibliography}

\bibliographystyle{elsarticle-num}
\bibliography{hpmc-paper}

\end{document}